# Voluntary rewards mediate the evolution of pool punishment for maintaining public goods in large populations

Tatsuya Sasaki[1,*], Satoshi Uchida[2], Xiaojie Chen[3]

[1]Faculty of Mathematics, University of Vienna, 1090 Vienna, Austria

[2]Research Center, RINRI Institute, 101-8385 Tokyo, Japan

[3]School of Mathematical Sciences, University of Electronic Science and Technology of China, 611731 Chengdu, China

[*]Corresponding to: tatsuya.sasaki@univie.ac.at

**Abstract:** Punishment is a popular tool when governing commons in situations where free riders would otherwise take over. It is well known that sanctioning systems, such as the police and courts, are costly and thus can suffer from those who free ride on other's efforts to maintain the sanctioning systems (second-order free riders). Previous game-theory studies showed that if populations are very large, pool punishment rarely emerges in public good games, even when participation is optional, because of second-order free riders. Here we show that a matching fund for rewarding cooperation leads to the emergence of pool punishment, despite the presence of second-order free riders. We demonstrate that reward funds can pave the way for a transition from a population of free riders to a population of pool punishers. A key factor in promoting the transition is also to reward those who contribute to pool punishment, yet not abstaining from participation. Reward funds eventually vanish in raising pool punishment, which is sustainable by punishing the second-order free riders. This suggests that considering the interdependence of reward and punishment may help to better understand the origins and transitions of social norms and institutions.

Cooperation is costly in the commons dilemma. The evolution of cooperation among nonrelatives with social learning has been a persistent issue approached interdisciplinary, as more than a biological issue[1,2]. Needless to say, those who free ride on the cooperation of others are better off than those who cooperate, unless structural changes are considered in the individual payoff. As is well known, various selective incentives, such as reward, punishment, or ostracism, have been used to modify payoff structures and curb human behaviors[3,4]. Thus far, theoretical and experimental studies have mostly focused on punishment[5,6], which can sustain a high level of cooperation in providing public goods[7-9].

The evolution of punishment, however, remains a challenging puzzle[6,10-12]. Punishing is costly. It is thus not an easy task to explore if and how costly punishment pays[11]. Previous studies on the evolution of punishment have also demonstrated that differences in the details of punishment schemes, in particular when a punisher's cost is incurred, can have a large effect[13,14]. One representative type that has been most studied is informal or *peer* punishment. Peer punishment is inductively modeled, being typically described as: because you wronged me (or someone), I will punish you. As such, peer punishment depends on the assessment of past behaviors[15].

Continuing costly punishment itself is another public good and thus peer punishment can pave the way for regression to the punishment of free riders through peer punishment (second-order punishment)[4,7,16]. In the same way, third-order punishment and so on are applied. This will result in an infinite regress of costly punishments. Or in situations in which punishment against contributors or retaliation is allowed, these acts can offset the payoff advantage of the existing prosocial punishers over free riders[17,18]. People afraid of antisocial and counter punishments thus might shift the responsibility for sanctioning to others[19].

Along this line of thinking, we turn to another representative type of punishment, formal or *pool* punishment. Pool punishment is a "preemption" system that is set in place before forming joint enterprises (i.e., a public good game) and without knowing if there is a free rider among the participants, and subsequently each participant is offered the opportunity to contribute to a fund for pool punishment[12,14,20-25]. Recent studies show that considering second-order punishment results in pool punishment becoming more effective than peer punishment for stabilizing a cooperative state and participants are more likely to prefer pool punishment over peer punishment[14,23,26,27]. In pool punishment, it is assumed that a centralized authority, once established, can exclusively control the incentives, so that it suppresses non-responsible punishment and excludes the possibility of free riders higher than second-order.

Apart from the issue of system stabilization, there still remains another issue relevant for the evolution of costly punishment: the emergence problem. Indeed, punishing right and left in large populations of free riders will require considerable effort and expense for pool punishers. Reflecting this, it is often explicitly assumed that pool punishment becomes active if at least a threshold number of players, more than one, contribute to it[28,29,30]. This means that in such large populations it is not easy to successfully start up costly punishment[31,32], even with considering punishment of second-order free riders[33] (Fig. 1a,b).

For the last decade, several attempts have tried to resolve the emergence problem. Most of the theoretical results have been based on assuming small, finite populations and analyzing those stochastic dynamics[13,24,34-36]. In addition, optional participation and mutual aid games (MAGs) have been considered as key factors in a resolution[14,22,25]. When participation in games is optional, players can simply escape a social trap of mutual defection[37,38]. MAGs are variants of public good games (PGGs). In PGGs the resulting benefits are shared equally among all members in the group. In MAGs it is not allowed to benefit from one's own contribution to the public goods provision[13,20]. That is, MAGs deal with excludable goods, not public goods, and combined with optional participation, are also two-fold exclusion. As such, previous studies have shed light on excludable good games in small populations.

Here we turn to pool reward[39,40], thereby we tackle the emergence of pool punishment in non-excludable good games in very large populations. We model a situation like a matching fund that usually arises for charity or common goods, in which contributors donate to a nonprofit source outside. Then the external source, enhancing the input, will make returns to a broader range of beneficiaries that include the contributors. Previous studies have investigated reward and punishment, often comparatively[41-45], and have also examined the selection or interplay of these incentives[46-52]. It is thus surprising that little is known about what happens if those who commit to pool punishment are promptly rewarded, rather than through iterated interactions or reputation.

Rewarding is costly. The pool reward being considered allows for receiving self-returns from one's own contribution as well as sharing in other's contribution without contributing, similar to PGGs. It follows that a pool reward can suffer from those who take a free ride on the reward-fund raising. From the viewpoint of its initiator, rather than punishing, rewarding can be less expensive and thus more efficiently stimulate cooperative behaviors[4,43,46,52]. Indeed, voluntary rewarding can be maintained even in public good games with second-order free riders[39,53]. It is thus predicted that a pool reward that also rewards volunteers to pool punishment will provide a foothold for the initially rare volunteers to proliferate, overcoming the emergence problem even without the assistance of the optional participation. We shall confirm this prediction by using the following game-theoretical model.

## Methods

**Evolutionary games for a public good and multi-strategy interactions.** We consider a well-mixed, infinitely large population. We assume that a player is more likely to adopt other player's strategy earning a higher payoff ("imitate better"). In the population this can be implemented by considering replicator dynamics[54,55]. We analyze the replicator dynamics for five strategies that consist of four types of participants in the PGGs: (i) cooperators (C) contribute to the PGG, but not to the incentive funds; (ii) defectors (D) do not contribute at all; (iii) punishers (P) contribute to the PGG and to the punishment fund; (iv) rewarders contribute to the PGG and to the reward fund (R); and (v) non-participants (N). We denote as $x_S$ and $P_S$ the relative frequency and expected payoff for each strategy $S =$ C, D, P, R, or N (thus, $0 \leq x_S \leq 1$ and $\sum_S x_S = 1$). The replicator dynamics for the five strategies are given by $\dot{x}_S = x_S(P_S - \bar{P})$, in which $\bar{P}$ describes the average payoff over the population, that is $\bar{P} = \sum_S x_S P_S$.

**Game procedure and parameters.** A group for the public good interaction consists of $n > 1$ members who are randomly chosen from the population. First, each of the members is offered an opportunity to participate in the PGG. If participating, then each participant will be subsequently offered distinct opportunities to contribute, to the reward, then the punishment, and finally PGG. Each contribution to the PGG, reward or punishment fund means an investment of $c_1$, $c_2$, $c_3 > 0$, respectively, at a cost to the contributor itself. In the PGG, the resulting benefits, multiplied by factors $r_1 > 1$, are equally shared by all participants, excluding N-players. To examine a previous, problematic situation in which C, D, and N coexist, in particular we assume that $2 < r_1 < n$[37]. In the reward fund, the resulting rewards, multiplied by intermediate factors $r_2$ with $1 < r_2 < n$, is shared, yet not always equally, among all of the contributors to the PGG (C-, R-, and P-players), excluding D- and N-players[39,40,43]. We assume weights $k_{RP}$, $k_{RR} \geq 0$ for the P- and R-players' share. In the punishment fund, non-contributors to the punishment (D-, C-, and R-players) incur fines. We assume that the fines are proportional to the contribution accumulated over all P-players[14,20], with proportionality factor $r_3 > 1$ and weights $k_{PC}$, $k_{PR} \geq 0$ for the C- and R-players' fines. Finally, the fifth type (v) non-participant is a loner that independently earns a small payoff $g > 0$. Hence, we have the individual payoff for an interaction, $f_S$, of each strategy $S = $ C, D, P, R, or N, as follows:

$$\begin{aligned}
f_C &= \frac{b_1(n_C + n_R + n_P + 1)}{n - n_N} - c_1 + \frac{b_2 n_R}{(n_C + 1) + k_{RR} n_R + k_{RP} n_P} - k_{PC} b_3 n_P, \\
f_D &= \frac{b_1(n_C + n_R + n_P)}{n - n_N} - b_3 n_P, \\
f_P &= \frac{b_1(n_C + n_R + n_P + 1)}{n - n_N} - c_1 + \frac{k_{RP} b_2 n_R}{n_C + k_{RR} n_R + k_{RP}(n_P + 1)} - c_3, \\
f_R &= \frac{b_1(n_C + n_R + n_P + 1)}{n - n_N} - c_1 + \frac{k_{RR} b_2 (n_R + 1)}{n_C + k_{RR}(n_R + 1) + k_{RP} n_P} - c_2 - k_{PR} b_3 n_P, \\
f_N &= g,
\end{aligned} \qquad (1)$$

in which $n_S$ denotes the number of S-player among $(n - 1)$ co-players, $b_1 = r_1 c_1$, $b_2 = r_2 c_2$, and $b_3 = r_3 c_3$. The expected payoff for each strategy is given by

$$P_S = \sum_{\substack{n_C + n_D + n_P + n_R + n_N = n-1 \\ 0 \leq n_C, n_D, n_P, n_R, n_N \leq n-1}} \frac{(n-1)!}{n_C! n_D! n_P! n_R! n_N!} x_C^{n_C} x_D^{n_D} x_P^{n_P} x_R^{n_R} x_N^{n_N} f_S \text{, in which}$$

$\dfrac{(n-1)!}{n_C! n_D! n_P! n_R! n_N!} x_C^{n_C} x_D^{n_D} x_P^{n_P} x_R^{n_R} x_N^{n_N}$ describes the probability of finding the specific $(n - 1)$ co-players which includes $n_S$ S-players ($S = $ C, D, P, R, and N).

Here, it has been assumed that there are participants of more than one, and if a participant is single, she or he acts as a non-participant and earns the same payoff $g$[31,37]. In the model we consider that the reward weight $k_{RP}$ and $k_{RR}$ describe an extra bonus for the one who contributed not only the PGG but also another public fund. Thus, $k_{RP}$ and $k_{RR}$ are supposed to be greater than 1. In the punishment weights, $k_{PC}$ and $k_{PR}$ are usually smaller than 1, denoting a discount factor for the one who did the second-order but not first-order free riding. For simplicity, we hereafter assume that $k_{RR}$ - 1 and $k_{PR}$ offset each other and in particular $k_{RR} = 1$ and $k_{PR} = 0$.

## Results

We, in terms of evolutionary game theory[54], show that voluntary rewarding for pool punishers can lead to a state in which all are P-players, no matter whether participation is compulsory or optional.

**Stability of a coercive society.** We start with analyzing local stability of the all-P state. In particular for the all-P state to be robust for the invasion of a rare C-player, we consider second-order punishment with $k_{PC}r_3 > 1/(n - 1)$, under which there is no temptation to switch to C when all play P, unless specifically stated otherwise. It is not difficult to also know from equation (1) under which conditions the all-P state is stable against the invasion of a rare D- or N-player. In the case of D this is when $c_1(1 - r_1/n) < c_3[(n - 1)r_3 - 1]$ holds, where the left and right sides describe the marginal costs for cooperating in PGGs and for being punished by $n - 1$ punishers, respectively. In the case of N the condition is that $g < c_1(r_1 - 1) - c_3$, where the right side means the payoff for the group of all P-players.

**Conditions of rock-scissors-paper cycles.** It is known that there can exist two kinds of periodic cycles among three strategies. It is clear that the last inequality above is also a sufficient condition that C dominates N. Considering also that N dominates D with $g > 0$ and that D dominates C with $r_1 < n$, it follows that when the PGG multiplication factor $r_1$ is greater than 2, C-, D-, and N-players alternatively become dominant in the population[37,38]. Otherwise, the population which consists of the three strategies will end up with a homogeneous state in which all play N[37]. We thus focus on PGGs with $r_1 > 2$ (and thus $n > 2$) in what follows. In addition, to hold such periodic oscillations among another triplet C-, D-, and R-players, it is necessary that $c_1(1 - r_1/n) < c_2(r_2 - 1)$ holds[39]. Based on these rock-scissors-paper-type cycles, we shall investigate the evolutionary dynamics for more than three strategies.

**With no reward, pool punishment never emerges** (Fig. 1a,b). We first consider combinations of C-, D-, P-players with or without N-players. We show that no P-players evolve if they are initially very rare, whatever the condition of participation. Let us start by compulsory participation (Fig. 1a). In a population which exclusively consists of P and D (or C), the replicator dynamics exhibit a bi-stable system: depending on the initial fraction of P-players in the population, the population can evolve either to a state of all P-players or a state of all D-players (or all C-players). By assumption D-players are always better off than C-players. Thus, for the three strategies, the dynamics exhibit bistability of the two homogeneous states for P-players or D-players (all-P state and all-D state). Next is in the case of optional participation (Fig. 1b). In competition among three strategies C, D, and N, it is supposed that the CDN face is filled with periodic closed orbits surrounding a center[37] (see Supplementary Fig. S1 for detailed phase portraits on the faces). For a coexisting state of C, D, and N within the CDN face, a rare, innovative P-player cannot invade, because the time average of the transversal growth rate (i.e., difference of the expected payoff of a rare P-payer and the average payoff over the population) for the rare P-player is negative per punishing cost $c_3$, which is the same as in the case of peer punishment[31]. Thus, in the given parameter settings, the dynamics exhibit bistability of the all-P state and periodic oscillations among C, D, and N (see Supplementary Fig. S2 for time series).

**With reward, pool punishment emerges for compulsory participation** (Figs. 1c and 2a). Replacing non-participation with a pool reward only leads to the similar dynamics on the corresponding CDR face, which is filled with periodic closed orbits surrounding a center[39] (Supplementary Fig. S1). It is obvious that the dynamics on the CDP face are unchanged. With an extra reward for P-players with $k_{RP} > 1$, the even rare P-player can be encouraged to invade the coexisting population on the CDR face. Numerical simulations show that the population state will typically come close to the DPR face, increasing in the fraction of P-players and decreasing in that of C-players. This is because of second-order punishment.

Among the three strategies of D, P, and R, the dynamics are repelling (Supplementary Fig. S1). As time goes by, the trajectories of population states will converge to the boundaries connecting the three homogenous states for D, P, and R. Considering that P-players are better off than R-players and R-players are better off than D-players, it is understood that the trajectories will be attracted to the all-P state, which again is robust for invasions of rare D- or C-players.

**Pool reward emerges for optional participation** (Fig. 3). To expand the applicable range of pool-reward, we also consider a case where participation is optional. It turns out that with sufficiently high degrees of the reward multiplication factor $r_2$, rare R-players can invade the CDN face, replacing N-players. The population state will eventually be attracted to a periodic orbit on the DNR face (see Supplementary Fig. S1 for detailed phase portraits on the faces). We remark that despite the fact that C-players exploit rewards by R-players, R-players can sprout in the presence of these second-order free riders. The successful invasion of a rare R-player deserves an example of the well-known Simpson's paradox[37,56,57] for second-order social dilemmas: in spite of the burden of costs for rewarding in each game, the rare R-player's payoff, when it is averaged over the whole population, will be better than the second-order free rider C-player's payoff[13,58]. This is in striking contrast to the former case in pool punishment (Fig. 1b). The DNR face, shared in Fig. 1c, is an "interface" to connect to the evolution of pool punishment and thus opens the door to the full course of the five strategies, as in what follows.

**With reward, pool punishment emerges for optional participation** (Fig. 2b). The initial state of the population almost exclusively consists of C-, D-, and N-players, and R- and P-players are given only at very small rates. The population first follows periodic oscillations among the resident three strategies. Similar to the last case, the initially rare R-players then start to gradually spread in the population, replacing N-players. The R-players then can take over almost all of the population. However, the all-P state finally arrives, substituting R-players. Without the intermediate sequence of a rise and fall of voluntary rewarding, we can only have continuous oscillations among C, D, and N.

**With no second-order punishment, pool punishment is unstable** (Supplementary Figs. S3 and S4). We explore effects of no punishment of C-players, who do not shoulder the punishment fee $c_3 > 0$. It is obvious that with no second-order punishment, C dominates P: switching the strategy from P to C is beneficial, whatever others do (see equation (1) with $k_{PC} = 0$). Therefore, the all-P state is unstable against the C's invasion and a small random shock will cause that a population of P-players will converge to a boundary state that completely excludes P-players. For compulsory participation with no reward, thus such a population will be eventually attracted by the all-D state (Supplementary Fig. S3a), and for optional participation with no reward, by cycles among C, D, and N (Supplementary Fig. S3b). With pool reward, however, the bonus weight for P-players $k_{RP} > 1$ can lead populations to temporally increase in P-players. The trajectories of population states then can converge to heteroclinic cycles, among C, D, R, and P for compulsory participation (Supplementary Figs. S3c and S4a). In particular, for optional participation the population will stay at an almost-all-N state for a long time on a heteroclinic cycle connecting the five homogeneous states for C, D, N, R, and P (Supplementary Fig. S4b).

We examine our main results for a certain range of the model parameters and initial conditions. Our main results that reward funds facilitate the emergence of costly pool punishment are robust against the various initial conditions, whether participation is optional or not (Supplementary Fig. S5). In particular we numerically explore the lower bound of the reward weight for punishers $k_{RP}$, with various settings of other parameters, $r_1$, $r_2$, $c_2$, and $c_3$ (Figs. 4b,d and Supplementary Fig. S6). In Figs. 1c and 2b we also investigate effects of (i) different group sizes $n$, (ii) different combinations of multiplication factors in PGGs and reward funds, $r_1$ and $r_2$ (Fig. 4a,c), (iii) nonlinearity of benefit production functions (Supplementary Text S1 and Fig. S7), and (iv) pool punishment which imposes constant fees. None of the variants (i)-(iv) qualitatively affect the main results.

## Discussion

Carrot or stick? This is a commonly used dichotomy in studies on selective incentives. Here we have focused on interdependence of reward and punishment. The evolution of costly punishment indeed will be promoted provided ample positive incentives that covers its net cost. In the case preferring costly punishment is a rational behavior. Thus, the core problem has been whether efforts to provide such rewards can endogenously evolve. Only a few studies have explored the evolution of a meta-norm that rewards peer punishers[59-62]. We have instead considered pool reward in *n*-person public good games, which can proliferate when rare even in the presence of second-order free riders. We examined a mediation effect of pool reward on overcoming the emergence problem of pool punishment. It turned out that considering pool reward leads to completing an evolutionary transition of societies in different equilibrium states, with norm deviators or norm followers. The latter state is protected by pool punishment.

Looking back to the real world, a law for an official subsidy or tax reduction to smoothly promote social changes (e.g., green cars and eco-friendly home) often includes its own expiration conditions. In our model, with achievement of a foothold for the evolution of pool punishment, the pool reward becomes evolutionarily retired. These mediation dynamics can be seen for variants of the model. For instance, rewarding mediation is applicable to nonlinear public good games in which the benefit production function has decreasing returns to scale[32]. This is also in threshold public good games in which a certain level of cooperation is required for producing public goods[40,63]. In either case, considering a sufficiently concave benefit function, the homogeneous state for cooperators turns into a stable state and even punishing free riders is redundant to maintain cooperation.

The essence of sustaining pool punishment is its prior commitment scheme followed by second-order punishment. Exploring if and how such a commitment system can emerge is out of the range of the model considered. Second-order punishment has been found to effectively prevent second-order free riders from eroding the voluntary sanctioning system[7,64,65]. In the case of peer punishment, it has also been reported that second-order punishment is not likely to be observed[62]. In contrast to this, pool punishment of second-order free riders is often conspicuously observed (i.e., against tax evaders). However, each individual is not supposed to transcendentally abide by the norm of pool punishment. In particular, in the very beginning when people never had concepts of pool punishment and thus there are also second-order free riders, how does a norm that assesses second-order free riders as bad emerge?[66] A better understanding of this could be relevant to the quest to understand the "roots of sanctioning institutions"[23]. As such, the fascinating origin of norm assessment for second-order pool punishment deserves further investigation.

Nowadays, various modern issues of commons, such as energy, natural environment, and climate change, are reaching every corner and covering all stages of human lives. As such, it appears that there is almost no time or space for people to opt out of both the corresponding dilemma situations and the related laws[34,67]. Results, based on compulsory participation but voluntary rewards, thus could be more applicable than previous theories with optional participation[14,34]. This implies an improved scenario to accomplish Garrett Hardin's recipe for the commons: mutual coercion mutually agreed upon[1]. In Isaiah Berlin's concept[68], optional participation (with "leaving loners alone"[36]) can be viewed as a *negative* liberty, freedom from interference in individual payoff by other players.

In contrast to this, voluntary rewards could be a *positive* liberty, freedom aimed at modifying the payoff of others. Recent studies have also shown that participants who enable an effect on one another through a majority vote prefer a coercive society with second-order pool punishment[27]. We have revealed that in a broad range of conditions with large populations, non-excludable public goods, or general benefit functions, only having optional participation is often not sufficient[32,67], but when combined with voluntary rewards, can be effective for establishing pool punishment. All in all, the results may suggest: through positive liberty, corrective coercion.

**Acknowledgments**

We thank Boyu Zhang for comments and suggestions. We also thank the Conference "Evolution of Cooperation", 8-10 April 2014, Sino-German Center for Research Promotion, Beijing, China, sponsored by National Natural Science Foundation of China (NSFC) and International Institute for Applied Systems Analysis (IIASA). T.S. was supported by the Foundational Questions in Evolutionary Biology Fund: RFP-12-21 and the Austrian Science Fund (FWF): P27018-G11.

**Author contributions**

T.S., S.U., and X.C. designed and wrote the paper. T.S. and S.U. performed research.

**Author information**

**Competing financial interests:** The authors declare no competing financial interests.

# Figures

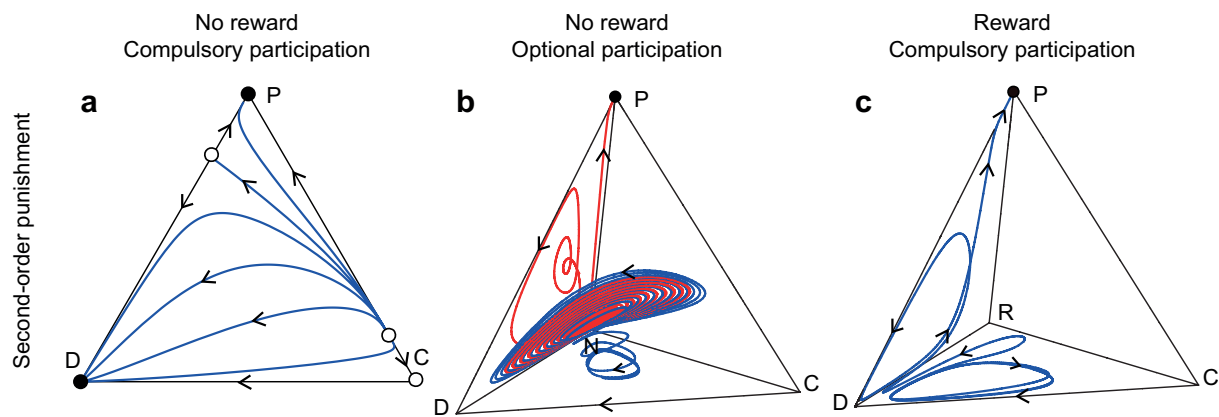

**Figure 1 | Evolution of pool punishment.** With no reward, (**a**) bistability of states with all P-players (P node) or all D-players (D node) for compulsory participation or (**b**) bistability of the P node or periodic oscillations among C, D, and N for optional participation. (**c**) Replace non-participant N with R. As in panel **b**, on the CDR face the population states oscillate along periodic closed orbits. In contrast to panels **a** and **b**, rare P-players, rewarded, can invade to the CDR face. Typically, the population state will converge to the DPR face, on which the dynamics is repelling. The trajectory then will come close to the edges connecting the three nodes D, P, and R, and finally attain the P node. Parameter values are: $n = 5$, $c_1 = 1$, $r_1 = 3$, $c_2 = 1$, $r_2 = 2$, $c_3 = 0.1$, $r_3 = 1.6$, $k_{RP} = 2$, $k_{PC} = 1$, and $g = 1$. The system includes second-order punishment. *Open* and *filled* circles denote, respectively, unstable and asymptotically stable equilibria.

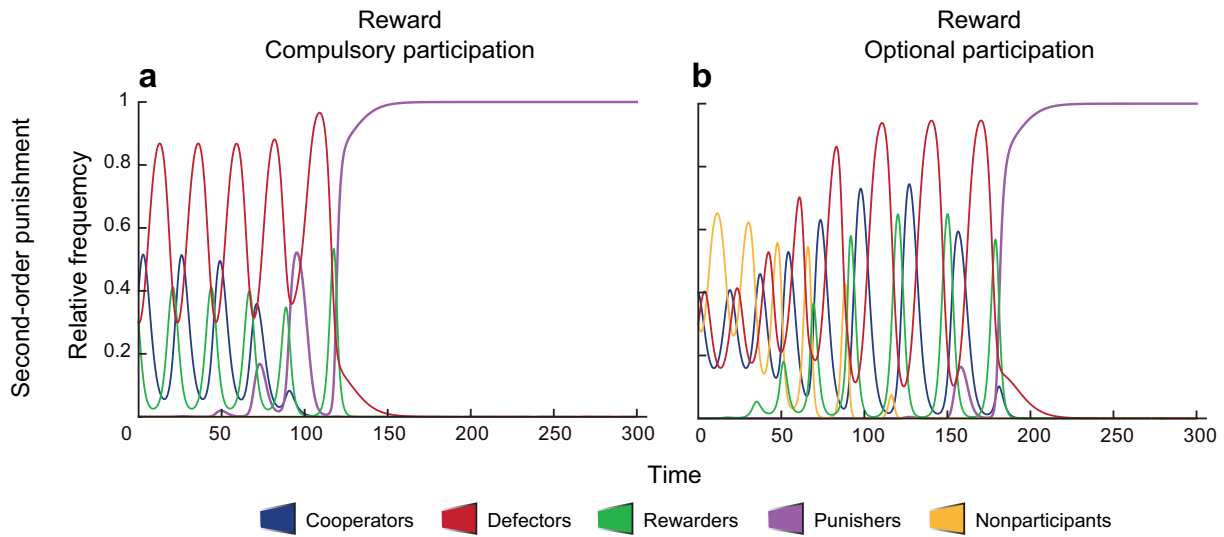

**Figure 2 | Via reward to punishment.** Time series of the frequencies of five strategies C (*blue*), D (*red*), P (*purple*), R (*green*), and N (*yellow*). (**a**) Participation is compulsory and thus N is excluded. Initially, C, D, and R are common and P is very rare. The population first follows periodic oscillations among C, D, and R. The rare P gradually invades and then takes over. (**b**) The initial state R and P are very rare. This population first follows periodic oscillations among C, D, and N. The rare R gradually invades and then takes over. The homogeneous state of P finally arrives, substituting the existence of R-players. Parameter values are as in Fig. 1. Initial conditions are: $(x_C, x_D, x_P, x_R, x_N)$ = (0.4, 0.2999, 0.0001, 0.3, 0) for panel **a**, or (0.4, 0.2998, 0.0001, 0.0001, 0.3) for panel **b**. The system includes second-order punishment.

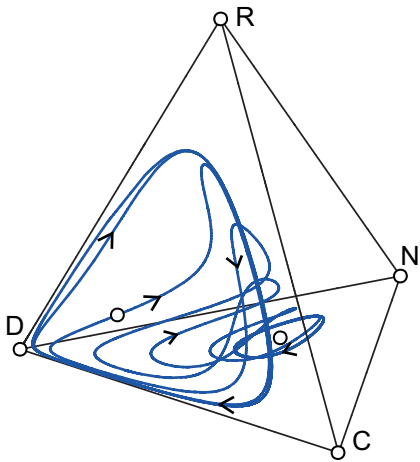

**Figure 3 | Via withdrawal to reward.** Participation is optional. Initially, C, D, and N are common and R is very rare. The population first follows periodic oscillations among C, D, and N. The rare R gradually invades the population, substituting the existence of N-players. The dynamics shift to periodic oscillations among C, D, and R. Parameter values are as in Fig. 1.

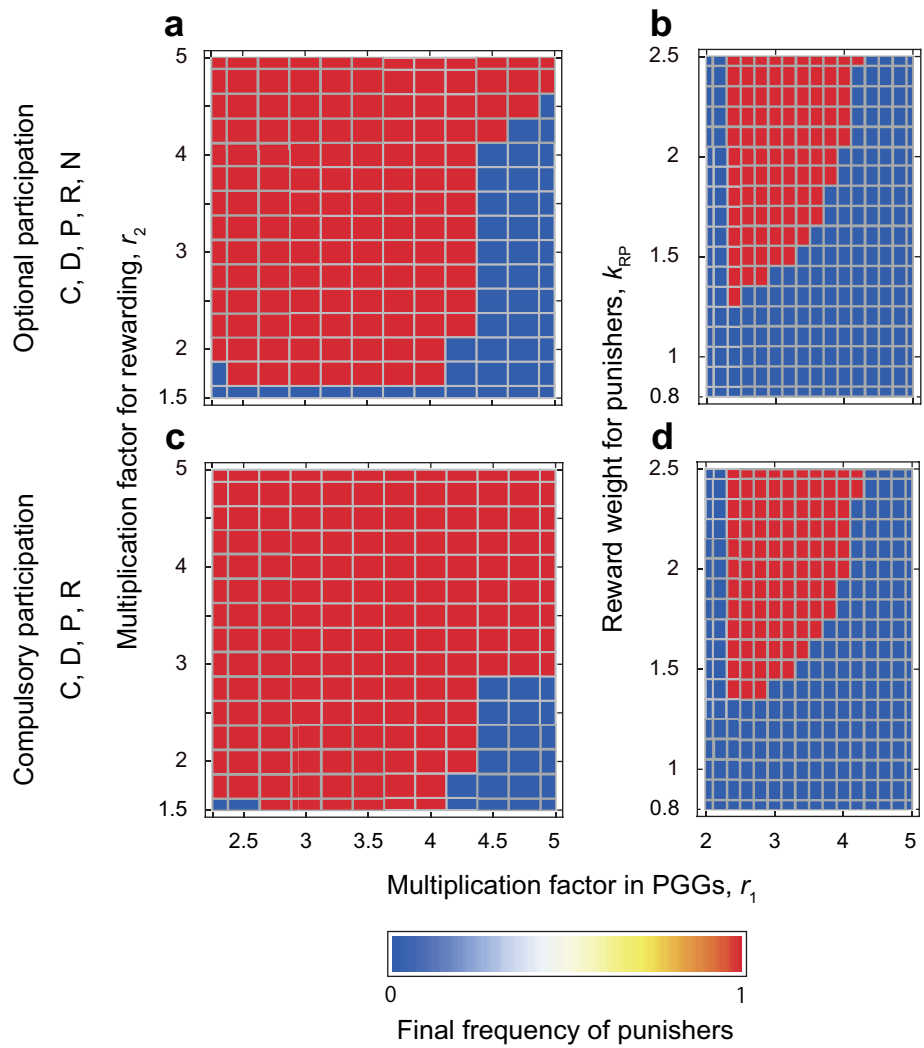

**Figure 4 | Effects of reward cost and weight on the evolution of pool punishment.** Initial conditions are: $(x_C, x_D, x_P, x_R, x_N) = (0.33, 0.338, 0.001, 0.001, 0.33)$ for panels **a** and **b**, and $(x_C, x_D, x_P, x_R) = (0.33, 0.339, 0.001, 0.33)$ for panels **c** and **d**. Other parameter values are as in Fig. 1.

**Supplementary Information for**

**Voluntary rewards mediate the evolution of pool punishment for maintaining public goods in large populations**

**This file include:**

Supplementary Text, S1
Supplementary Figures, S1-S7
Legends for supporting figures, S1-S7

**Text S1 Model variants**

**Quadratic benefit functions.** We can investigate some variants in benefit functions, which in the main text have been linear proportionally to $c_1 r_1$. Here we extensively examine quadratic functions for the provision of the total benefit, as follows: $B(X) = c_1(r_{12}X^2 + r_1 X)$, in which $X$ denotes the number of contributors to the public good: $X = n_C + n_P + n_R$. The first term in the right side of equation (1) is replaced with $B(X)/(n - n_N)$. In Supplementary Fig. S7 it turns out that concavity with $r_{12} < 0$ or convexity with $r_{12} > 0$ can lead the center point $Q_{RC}$ on the CDR face to turn, respectively, into an attractor or a repeller (Supplementary Fig. S7).

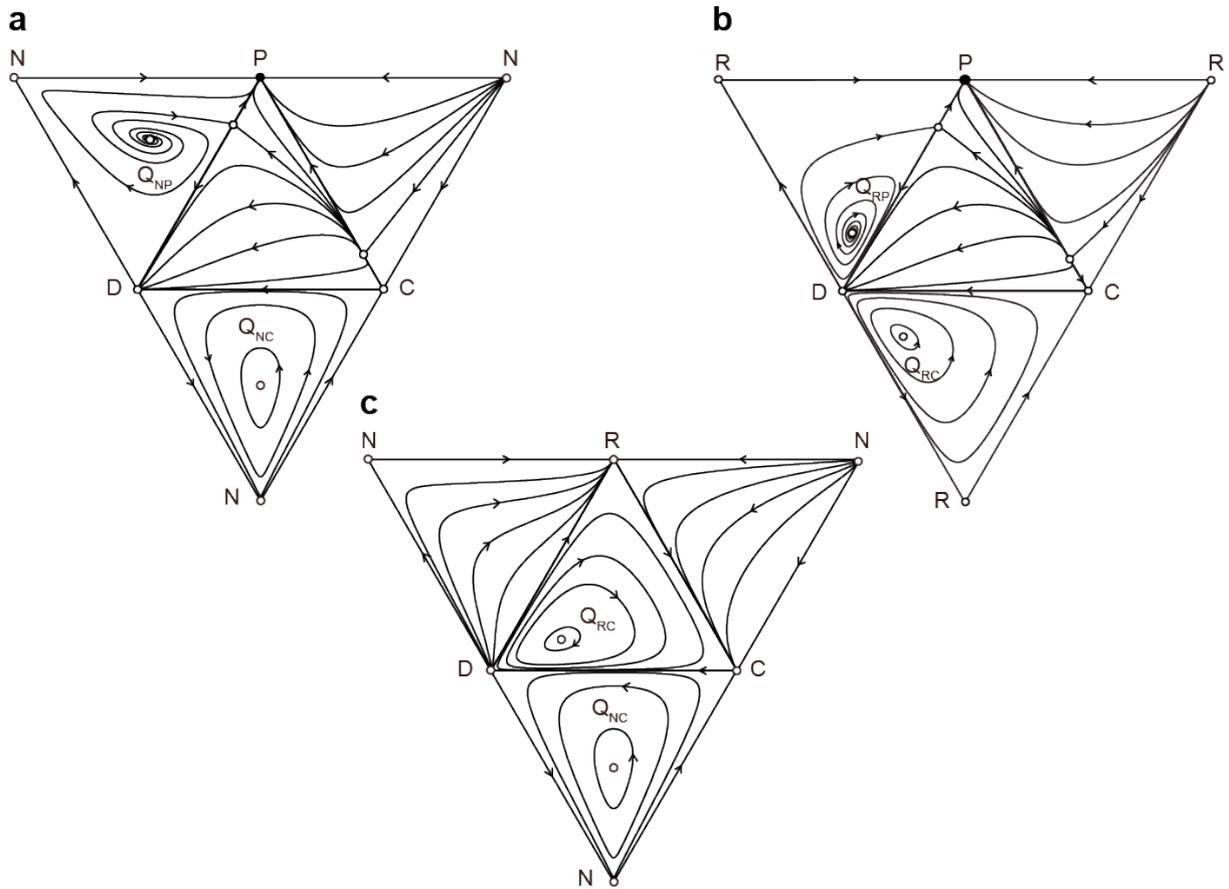

**Figure S1 | Replicator dynamics on boundaries.** Public good games with (**a**) optional participation and pool punishment, (**b**) compulsory participation and pool reward, or (**c**) optional participation and pool reward. Each simplex component describes a phase portrait associated with the replicator dynamics for three strategies displayed at the corners. *Open* and *filled* circles denote unstable and asymptotically stable equilibria, respectively. The corresponding 3-D phase portrait for four strategies is given in Figs. 1b, 1c, or 3. We note that for panels **a** and **b** the evolutionary dynamics are qualitatively similar on the faces. Indeed, the CDP face is common. And, the replicator dynamics on CDN in panel **a** and CDR in panel **b** lead to cyclical oscillations, on CPN in panel **a** and CPR in panel **b**, to bistability for P and C, and on DPN in panel **a** and DPR in panel **b**, to a repeller on the face and convergence to P. Despite this fact, interestingly, the interior dynamics, demonstrated in Figs. 1b and 1c, are strikingly contrast to each other in the aspect of bistability. Parameter values are: $n = 5$, $c_1 = 1$, $r_1 = 3$, $c_2 = 1$, $r_2 = 2$, $c_3 = 0.1$, $r_3 = 1.6$, $k_{RP} = 2$, $k_{PC} = 1$, and $g = 1$, as in Fig. 1.

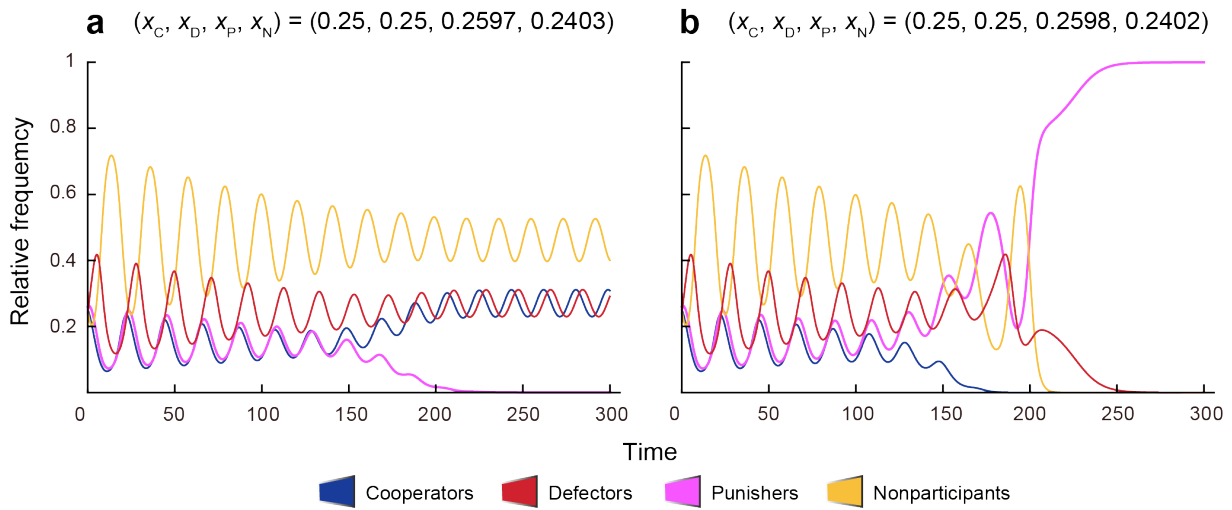

**Figure S2 | Sensitive responses to initial conditions in optional public good games with pool punishment.** Time series of the frequencies of four strategies, C (*blue*), D (*red*), P (*purple*), and N (*yellow*), corresponding to Fig. 1b. (**a**) The population continues periodically oscillating while the frequency of P gradually decreases and finally vanishes. (**b**) The initial frequencies of P and N are only 0.0001 more are less than those in panel **a**. The population starts with similarly oscillating, then instead of P-players, C-players first vanish, followed by extinction of N- and D-players. This leads to attaining the all-P state (which is stable by second-order punishment). Parameter values are as in Fig. 1. Initial conditions are: ($x_C$, $x_D$, $x_P$, $x_N$) = (0.25, 0.25, 0.2597, 0.2403) for panel **a** or (0.25, 0.25, 0.2598, 0.2402) for panel **b**. The system has second-order punishment.

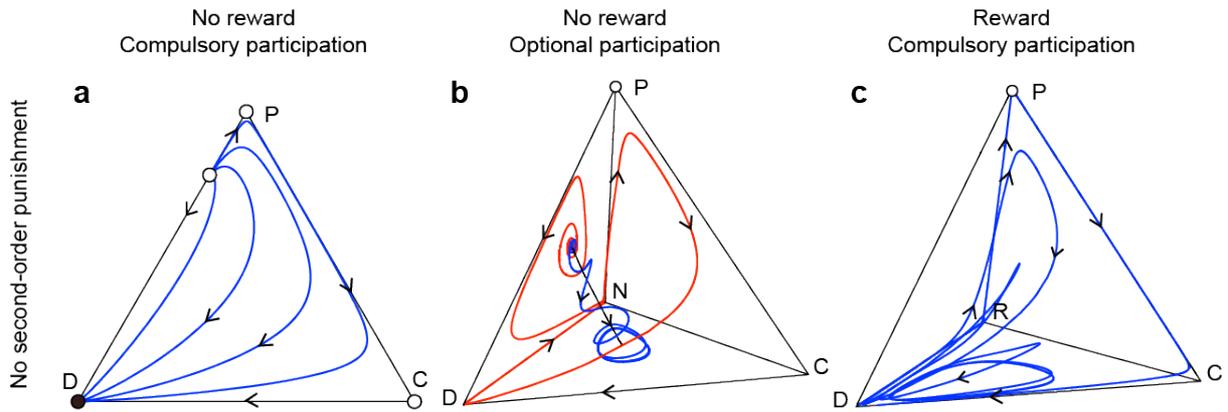

**Figure S3 | Evolution of pool punishment without second-order punishment.** The P node is no longer stable. (**a**, **b**) With no reward. The population can converge to the D node in compulsory participation in panel **a** or the CDN face in optional participation in panel **b**. (**c**) With reward. The population can converge to the heteroclinic cycle connecting the nodes C, D, R, and P in compulsory participation. Parameter values are: $n = 5$, $c_1 = 1$, $r_1 = 3$, $c_2 = 1$, $r_2 = 2$, $c_3 = 0.1$, $r_3 = 1.6$, $k_{RP} = 2$, $k_{PC} = 1$, and $g = 1$. *Open* and *filled* circles denote unstable and asymptotically stable equilibria, respectively.

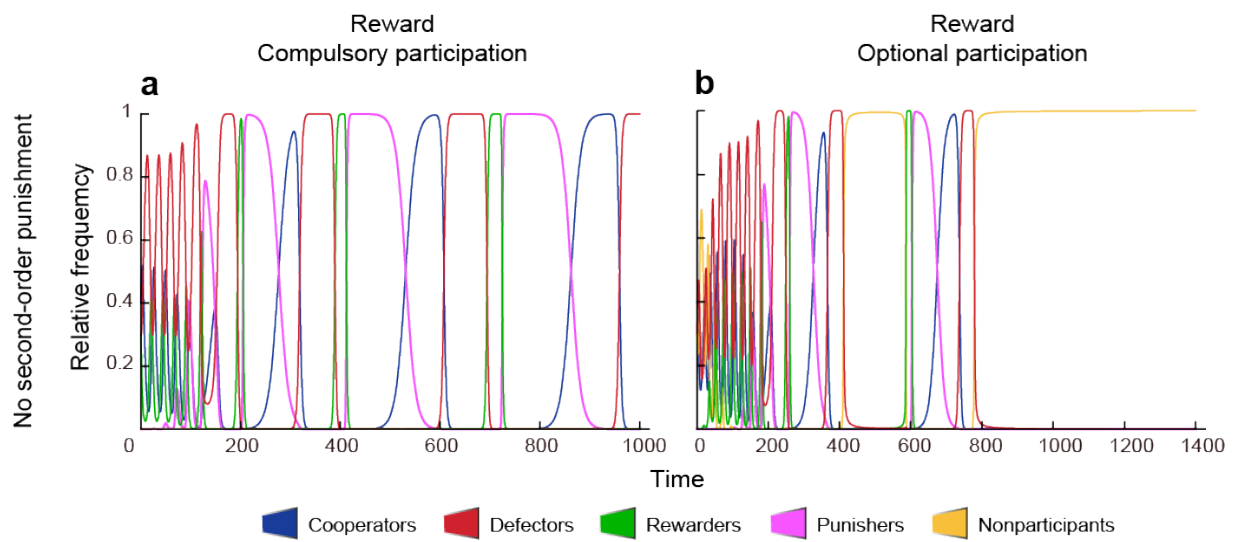

**Figure S4 | Cycles without second-order punishment.** Time series of the frequencies of five strategies C (*blue*), D (*red*), P (*purple*), R (*green*), and N (*yellow*). The system has no second-order punishment. The homogeneous state of P is no longer stable. Instead the population converges to heteroclinic cycles; in particular with optional participation in panel **a**, the population will stay in the homogeneous state of N for a long time. Parameter values are as in Fig. 1. Initial conditions are: $(x_C, x_D, x_P, x_R, x_N) = (0.4, 0.2999, 0.0001, 0.3, 0)$ for panel **a**, or $(0.4, 0.2998, 0.0001, 0.0001, 0.3)$ for panel **b**.

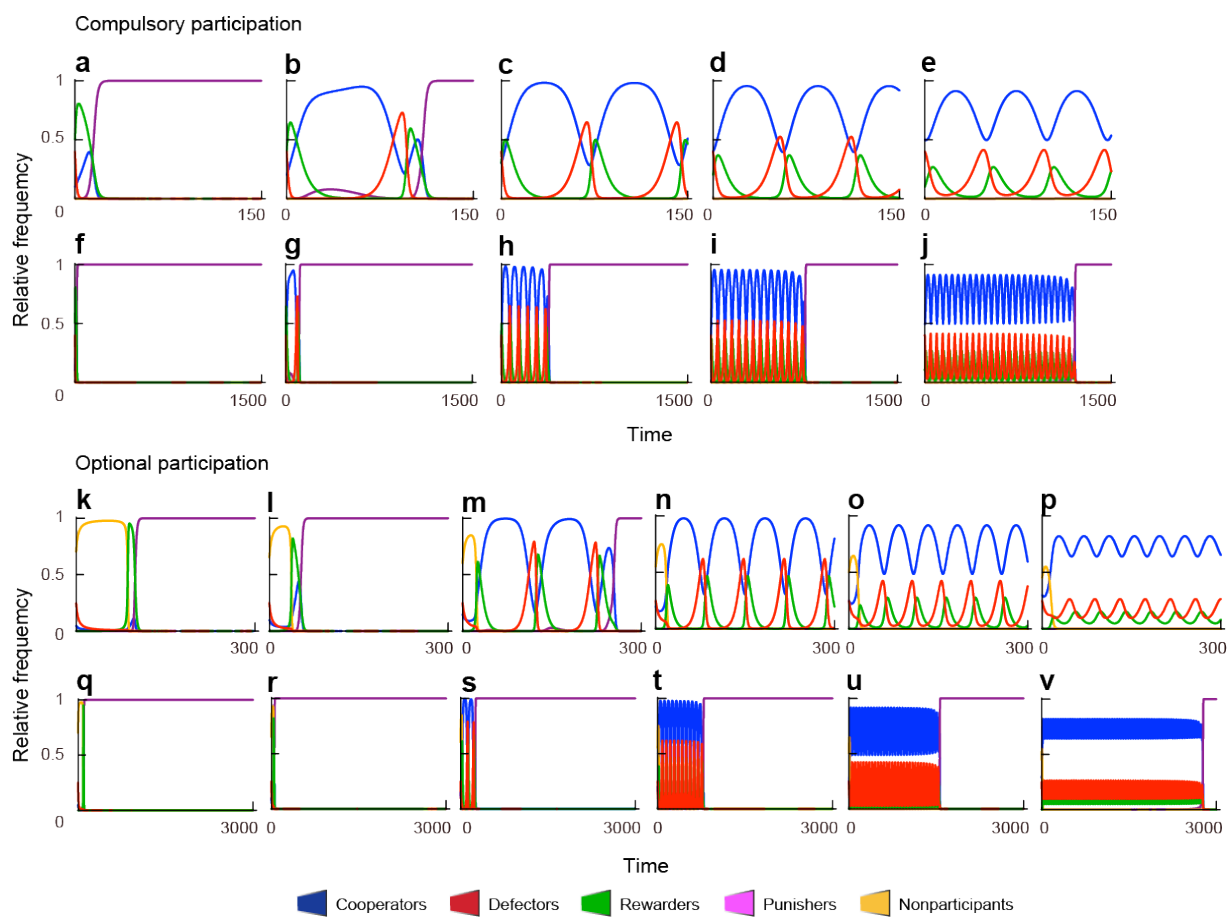

**Figure S5 | Effects of initial conditions in compulsory/optional public good games with pool reward and punishment.** Time series of the frequencies of (**a-j**) four strategies, C (*blue*), D (*red*), R (*green*), and P (*purple*), (**k-v**) five strategies, C, D, R, P, and N (*yellow*). Parameter values are as in Fig. 1. Initial conditions are: $(x_C, x_D, x_P, x_R, x_N) = (0.1, 0.39999, 0.00001, 0.5, 0)$ for panels **a** and **f**, $(0.2, 0.39999, 0.00001, 0.4, 0)$ for panels **b** and **g**, $(0.3, 0.39999, 0.00001, 0.3, 0)$ for panels **c** and **h**, $(0.4, 0.39999, 0.00001, 0.2, 0)$ for panels **d** and **i**, $(0.5, 0.39999, 0.00001, 0.1, 0)$ for panels **e** and **j**, $(0.05, 0.24998, 0.00001, 0.00001, 0.7)$ for panels **k** and **q**, $(0.1, 0.24998, 0.00001, 0.00001, 0.65)$ for panels **l** and **r**, $(0.15, 0.24998, 0.00001, 0.00001, 0.6)$ for panels **m** and **s**, $(0.2, 0.24998, 0.00001, 0.00001, 0.55)$ for panels **n** and **t**, $(0.25, 0.24998, 0.00001, 0.00001, 0.5)$ for panels **o** and **u**, or $(0.3, 0.24998, 0.00001, 0.00001, 0.45)$ for panels **p** and **v**.

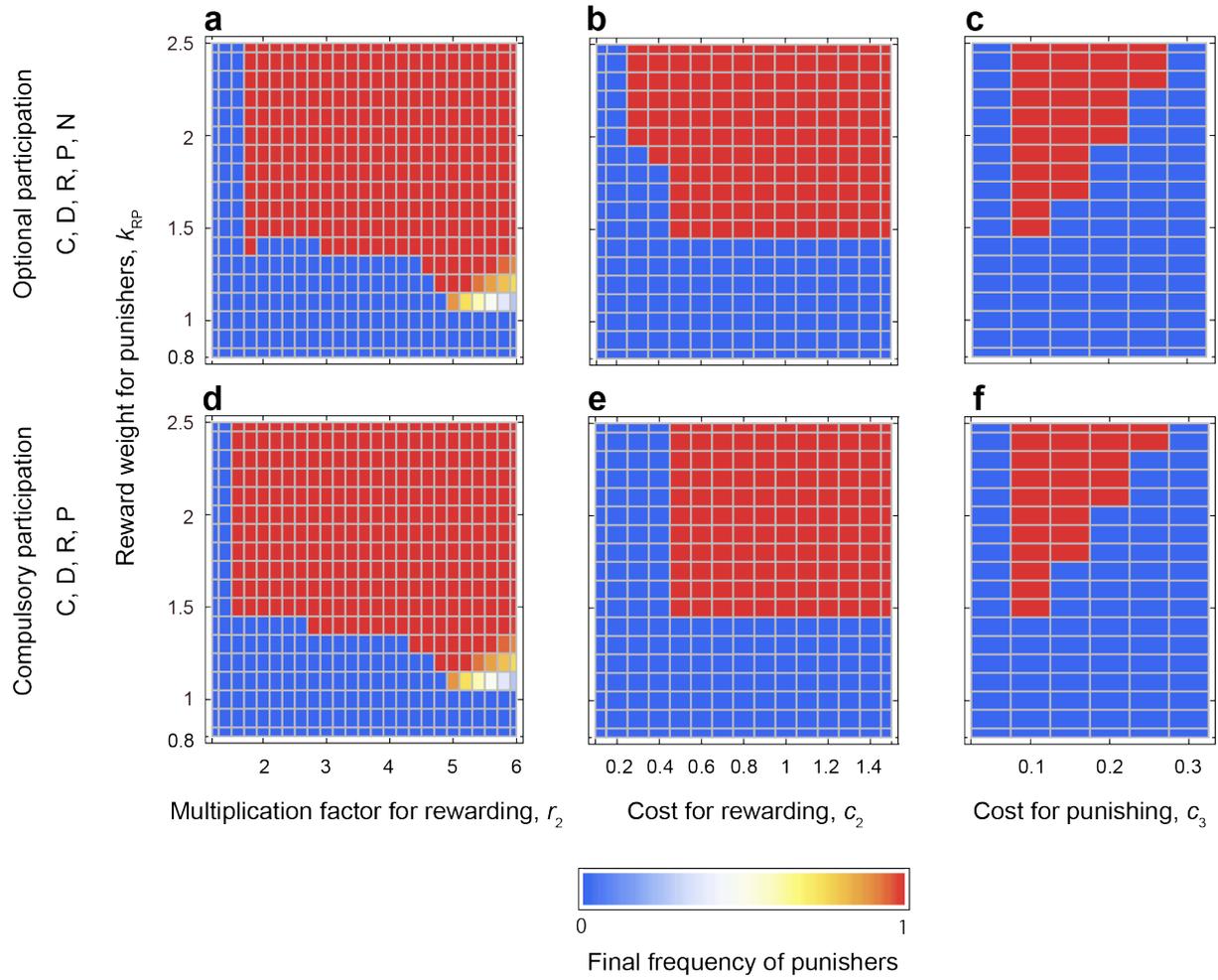

**Figure S6 | Parameter analyses for different incentive multipliers, costs, and weights.** Initial conditions are: $(x_C, x_D, x_P, x_R, x_N) = (0.33, 0.338, 0.001, 0.001, 0.33)$ for panel **a**, $(0.4, 0.298, 0.001, 0.001, 0.3)$ panels **b** and **c**, $(0.33, 0.339, 0.33, 0.001, 0)$ for panel **d**, $(0.4, 0.299, 0.001, 0.3, 0)$ for panels **e** and **f**. Other parameter values are as in Fig. 1.

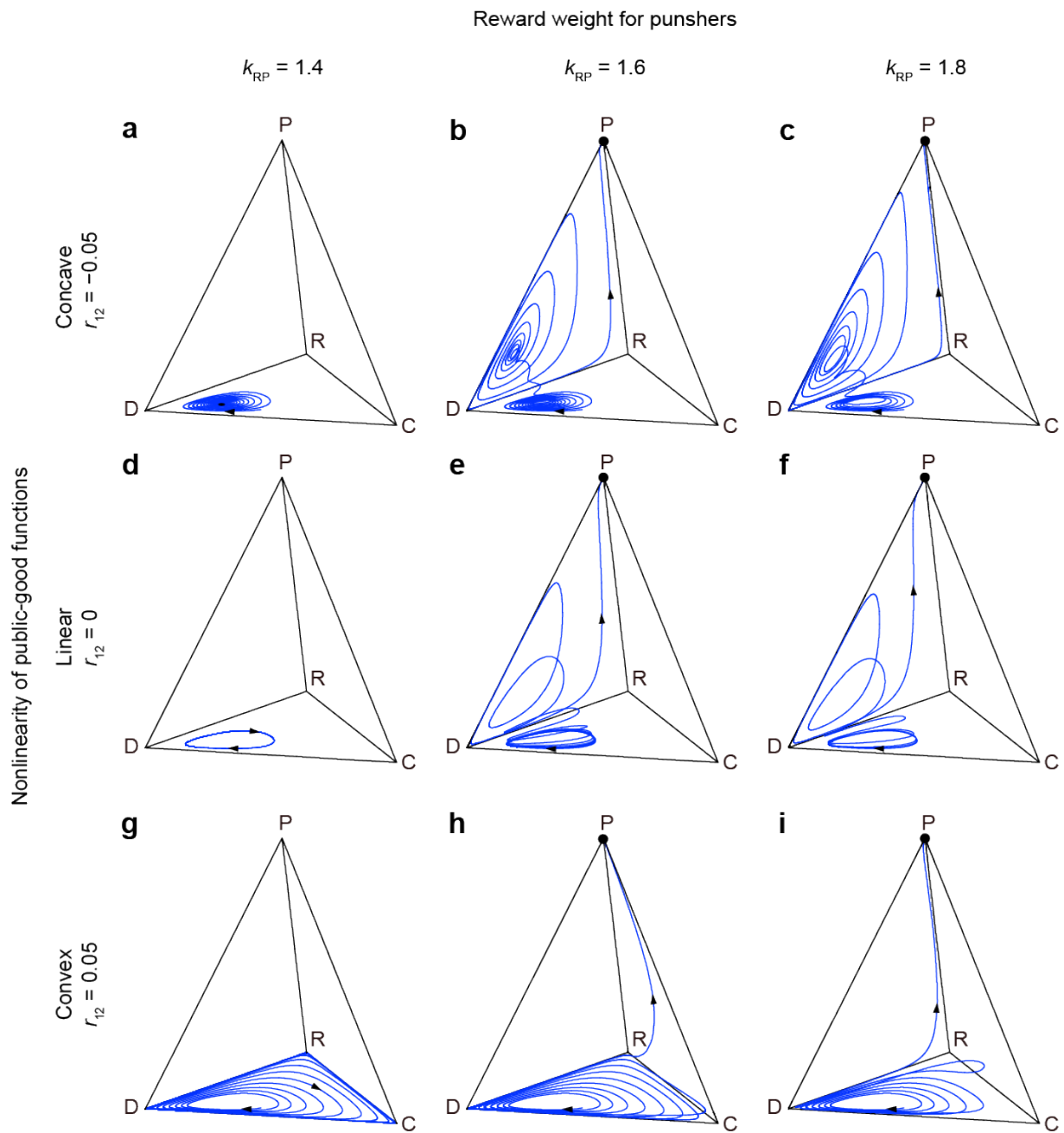

**Figure S7-1 | Responses to nonlinearity of benefit functions in compulsory public good games with pool reward and punishment.** The corresponding time series of the relative frequencies of four strategies, C, D, R, and P, are in Supplementary Fig. S7-2. (**a-c**) Concave benefit functions lead to an attractor on the CDR face. (**d-f**) Linear benefit functions lead to periodic closed orbits on the CDR face. (**g-i**) Convex benefit functions lead to a repeller on

the CDR face. In spite of those differences in nonlinearity, the population state can eventually converge to the all-P state, as the reward weight for punishers $k_{RP}$ are sufficiently large. Parameter values are as in Fig. 1. Initial conditions are: $(x_C, x_D, x_R, x_P)$ = (0.4, 0.49999, 0.1, 0.00001).

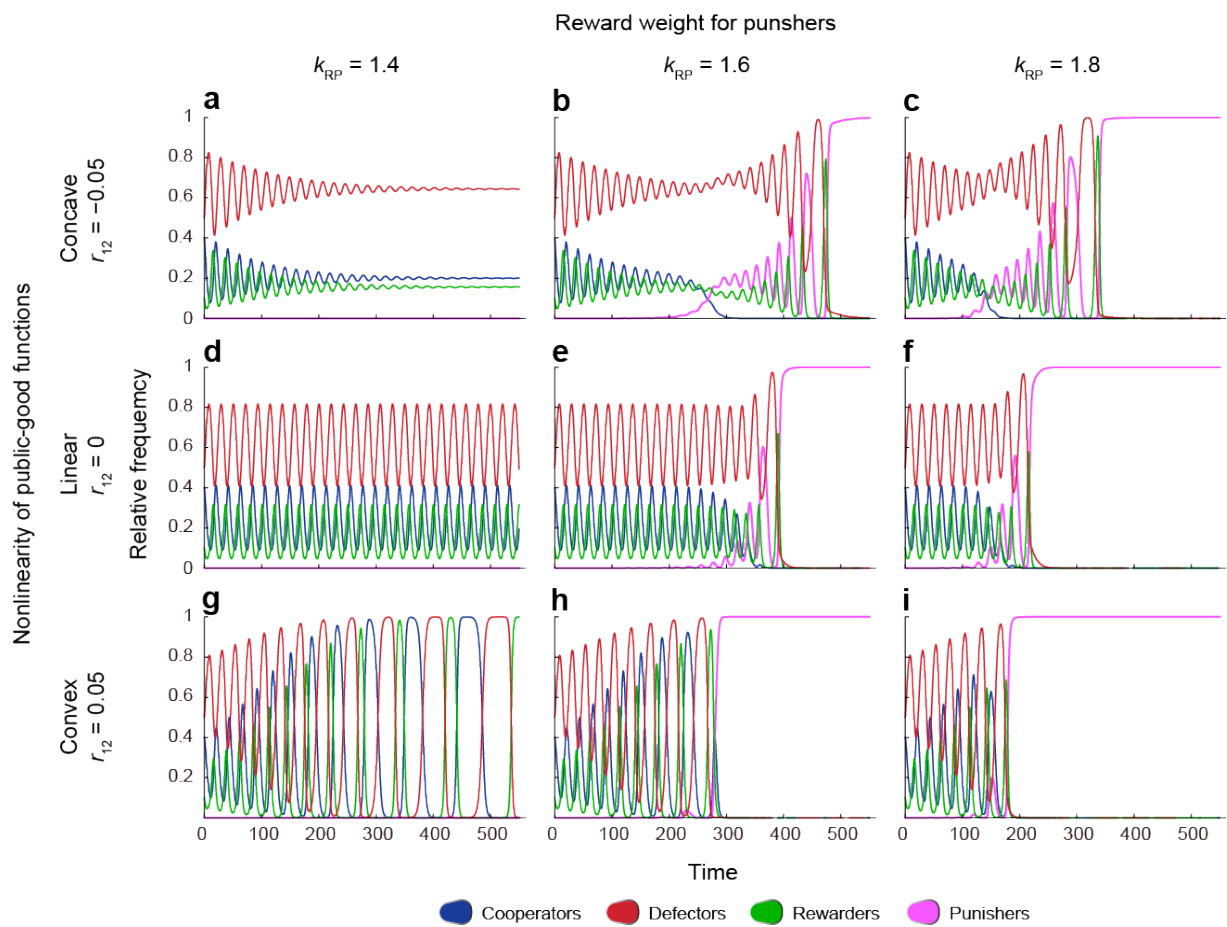

**Figure S7-2 | Responses to nonlinearity of benefit functions in compulsory public good games with pool reward and punishment.** Time series of the frequencies of four strategies, C (*blue*), D (*red*), R (*green*), and P (*purple*), corresponding to Supplementary Fig. S7-1.